\documentstyle[spro,psfig]{article}

\def\be{\begin{equation}}
\def\ee{\end{equation}}
\def\bea{\begin{eqnarray}}
\def\eea{\end{eqnarray}}

\begin{document}

\title{GIANT RESONANCES IN COULOMB EXCITATIONS OF RELATIVISTIC IONS
}

\author{ M.S. Hussein, A.F.R. de Toledo Piza, 
O.K. Vorov
}

\address{
Instituto de Fisica, 
Universidade de Sao Paulo \\
Caixa Postal 66318,  05315-970,  \\
Sao Paulo, SP, Brasil 
}


\maketitle\abstracts{
We propose a soluble model to incorporate the nonlinear effects in the
transition probabilities of the multiphonon Giant Dipole Resonances
based on the SU(1,1) algebra. 
Analytical expressions for the
multi-phonon transition probabilities are derived. 
For reasonably small magnitude of nonlinearity $x\simeq 0.1-0.2$
enhancement factor for the
Double Giant Resonance excitation probabilities and the cross sections
reaches values $1.3-2$ compatible with experimental data. 
}

\section{Introduction}

Coulomb Excitation in collisions of relativistic ions is one of
the most promising methods in modern nuclear physics 
\cite{EMLING,BCHTP,bbabs,CH-FR,HBCH}. 
One of the most interesting applications of 
this method to studies of nuclear structure is the possibility to 
observe and study the multi phonon Giant Resonances \cite{EMLING}.
In particular, the double Dipole Giant Resonances (DGDR) have been 
observed in a number of nuclei 
\cite{SCH-EXP-Xe,R-EXP-Pb,AU-EXP-Au}. 
The ``bulk properties''
of the one- and two-phonon GDR are now partly understood \cite{EMLING}
and they are in a reasonable agreement with the theoretical picture
based on the concept of GDR-phonons as almost harmonic quantized vibrations.

Despite that, there is a persisting discrepancy between the theory and the
data, observed in various experiments 
\cite{SCH-EXP-Xe,R-EXP-Pb,AU-EXP-Au,B-208-Pb-1.33,B-208-Pb-LOW} that still 
remains to be understood: the double GDR excitation cross sections 
are found enhanced by factor $1.3-2$ with respect to the predictions
of the harmonic phonon picture 
\cite{EMLING},\cite{bbabs},\cite{AW65},\cite{WA79}.
This discrepancy,
which almost disappears at high bombarding energy,
has  attracted much attention in current literature 
\cite{CH-FR},
\cite{CRHP94,Norbury-Baur,hot-phonon1,hot-phonon2,BZ93},
\cite{vccal,vca2,B-D};
among the approaches to resolve the problem are the higher order
perturbation theory treatment \cite{BZ93},  and studies of 
anharmonic/nonlinear aspects of GDR dynamics \cite{CH-FR},
\cite{vccal,vca2},\cite{B-D}.
Recently, the concept of hot phonons \cite{hot-phonon1},\cite{hot-phonon2}
within Brink-Axel mechanism was proposed that provides microscopic 
explanation of the effect. 
These seemingly orthogonal explanations deserve clarification which 
we try to supply here.

The purpose of this work is to examine, within a a soluble model 
the role of the
nonlinear effects 
on the transition amplitudes
that connect the multiphonon states in a heavy-ion Coulomb excitation 
process. Most studies of anharmonic corrections \cite{vccal,vca2,B-D}
concentrated on their 
effect in the spectrum 
\cite{CCG,Bertsch-Feldmeier}.

Within our model, the nonlinear effects are described by a single 
parameter, and the model contains the harmonic model as its limiting case
when the nonlinearity goes to zero. 
We obtain analytical expressions for the probabilities of excitation
of multiphonon states which substitute the Poisson formula of the 
harmonic phonon theory. 
For the reasonable values of the nonlinearity, the present model
is able to describe the observed enhancement of the double GDR
cross sections quoted above.
 
Having in mind to show how analytical results follows from the
nonlinear model, and to explain how the model works, we restrict 
ourselves
here to its simplest version (transverse approximation, or SU(1,1)
dynamics) and keep numerics up to minimum level.
We postpone till further publications detailed numerical analysis
and comparison with the data. Microscopic origins of the nonlinear 
effects (which are considered here phenomenologically) are also
beyond the scope of our discussion.

\section{Excitation of multi-phonon GDR: Coupled-channels problem }

We work in a semiclassical approach \cite{AW65} to the
coupled-channels problem, i.e.,  
the projectile-target relative
motion is approximated by a classical trajectory and the
excitation of the Giant Resonances is treated 
quantum mechanically \cite{CRHP94}.
The use of this method is justified due to the small wavelenghts
associated with the relative motion 
in relativistic heavy ion collisions. 
The separation coordinate is treated as a classical time 
dependent variable, and the projectile motion 
is assumed to be a straight line \cite{BCHTP}.

The intrinsic
dynamics of excited nucleus is governed by a time dependent quantum 
Hamiltonian
(see Refs. \cite{AW65},\cite{WA79}).
The intrinsic state $\vert \psi(t)>$ of excited nucleus 
is the solution of the time dependent Schr\"odinger equation
\begin{equation} \label{SCHR}
i {\partial \vert \psi(t)> \over \partial t} =
\left[ H_0 + V (t) ) \right]\ \vert \psi(t)> ,
\quad
\vert \psi (t)> = \sum_{N=0}\ a_N(t)\ \vert N>\ \exp \Big(-i
E_N t \Big),
\end{equation}
where  $H_0$ is the intrinsic Hamiltonian and $V$ is the
channel-coupling interaction. We let 
$\hbar=1$, $c=1$.
The standard 
coupled-channel problem is to find the expansion amplitudes $a_N(t)$ 
in the wave packet $|\psi \rangle$
as functions of impact parameter $b$
where $E_N$ is the energy of the state $\vert N>$ specified by the
numbers of excited GDR phonons $N$. 
We assume the colliding nuclei to be in their ground states 
before the collision. 
The amplitudes obey the initial condition $a_N(t\rightarrow -\infty)
= \delta_{N,0}$ and they tend to constant values as
$t\rightarrow \pm \infty$ (the interaction $V(t)$ dies out at 
$t\rightarrow \pm \infty$). 
The excitation probability of an intrinsic
state $\vert N> $ in a collision with impact parameter $b$ 
and the 
total cross section for excitation of the state $\vert N>$ (given by the
integral over the imact parameter) are
\begin{equation} \label{Pn}
W_N(b) = \vert a_N(\infty)\vert ^2 , \qquad
\sigma_N= 2 \pi \int\limits_{b_{gr}}^{\infty} b W_N(b)  db  
\end{equation}
(we use the grazing value $b_{gr}= 1.2 (A_{exc}^{1/3}+A_{sp}^{1/3})$ as the
lower limit). Hereafter, the labels $exc$ ($sp$) refer 
to the excited (spectator) partner in a colliding projectile-target pair. 
We neglect the here nuclear contribution 
\cite{HRBHPT} to the excitation process.

It is convenient to treat
the coupled channel equations (\ref{SCHR})
in terms of the unitary operator $U_I$ (the interaction picture):
\begin{equation} \label{U}
i \frac{d} {d t} U_I(t) = V_I(t)  U_I(t) , 
\qquad V_I(t)= e^{iH_0 t} V(t) e^{-iH_0 t}, 
\qquad U_I(t=-\infty) = I,
\end{equation}
where the time-dependent Hamiltonian $H(t)=H_0+V(t)$ that acts in the intrinsic
multi-GDR states is given by 
$H_0 = \omega \hat{N}_d, \quad \hat{N}_d \equiv \sum\limits_{m}d^+_{m} d_{m}$ and
\begin{eqnarray} \label{BASIC}
V(t) = v_1(t) [ (E1_{-1})^{\dagger} - (E1_{+1})^{\dagger} ] +
v_0(t)   (E1_{0})^{\dagger} + Herm.Conj.  
\end{eqnarray}
where $E1_{m}^{\dagger}$ and $E1_{m}$ 
are the dimensionless operators acting in the
internal space of the multi-GDR states. 
The functions $v$ are given in \cite{BCHTP}, 
e.g.,
\begin{eqnarray} \label{v1}
v_1(t) = \frac{w}{[1 + (\frac{\gamma v}{b}t)^2]^{3/2}}, 
\quad w=   \rho \frac{Z_{sp} e^2 \gamma}{2 b^2} 
\sqrt{ \frac{ N_{exc} Z_{exc} }{A_{exc}^{2/3} m_N \cdot 80 MeV} }.
\end{eqnarray}
Here, $m_N$ and $e$ are the proton mass and charge,
$Z$, $N$ and $A$ denote the nuclear charge,
the neutron number and the mass number of the colliding partners,
$\gamma=(1-v^2)^{-1/2}$ is relativistic factor, $v$ is the velocity and
the parameter $\rho$ is the deal of the strength absorbed by the
collective motion (usually assumed to be close to unity) \cite{EMLING}.

In the harmonic approximation, 
the operators $E1_{m}^{\dagger}, E1_{m}$
are given by the GDR phonon creation 
and destruction operators of corresponding angular momentum projection $m$, 
$E1_{m}^{\dagger}=d^{\dagger}_m$.
This model of ``ideal bosons'' coupled linearly to the Coulomb field
has well known exact nonperturbative solution
(see, e.g. \cite{WA79}) 
for the excitation probabilities 
\begin{eqnarray} \label{POISSON}
W_N = e^{-|\alpha^{harm}|^2} \frac{|\alpha^{harm}|^{2N} }{N!},
\qquad \qquad \qquad 
\nonumber\\
\qquad 
|\alpha^{harm}|^2 
= \sum_{m=0,\pm 1} |\alpha^{harm}_m|^2 = 
2 |\alpha^{harm}_1|^2 + |\alpha^{harm}_0|^2,
\end{eqnarray}
i.e., the Poisson formula
where the amplitudes $\alpha^{harm}_m$ are expressed in terms of the 
modified Bessel functions. At the colliding energies sufficiently high,
the longitudinal contribution ($\propto |\alpha^{harm}_0|^2$) is
suppressed by a factor proportional to $\gamma^{-2}$,  
see, e.g., \cite{bbabs}.
In the following, we will work in this ``transverse approximation''
dropping the longitudinal term 
(the results are still 
qualitatively valid at lower energies). 

\section{Nonlinear Model}
Our idea is to keep the spectrum of GDR system harmonic with the
Hamiltonian $H_0 = \omega N$. That is supported by the systematics
of the observed DGDR energies, $E_2$, which yields 
$E_2 \simeq (1.75-2) \omega$ \cite{EMLING}.
The conclusion on the weak anharmonicity in the spectrum follows also
from theoretical considerations \cite{CCG},\cite{Bertsch-Feldmeier}.

The transition operators $E1^{\dagger}, E1$ that couple intrinsic 
motion to the Coulomb field can however include 
nonlinear effects: one can use the boson-expansion-like expression 
for them 
\begin{equation} \label{EXPANSION}
E1_{m}^{\dagger}=d^{\dagger}_m +  
x d^{\dagger}_{m_1}
d^{\dagger}_{m_2} d_{m_3} + 
x_1 d^{\dagger}_{m'_1} d^{\dagger}_{m'_2} d^{\dagger}_{m'_3} + ...
\end{equation}
where the parameters $x_i$ determines the strengths of the
nonlinear contributions.
In particular, this could be a result of 
nonlinearities (anharmonicities) in the phonon Hamiltonian obtained 
in higher orders of perturbation theory.
A reasonable way to take into account these nonlinear effects that 
we adopt in this work 
is to keep in the expansion (\ref{EXPANSION}) infinite series of type 
\begin{eqnarray} \label{E1}
E1_{m}^{\dagger}=d^{\dagger}_m +  x\sum_{m_1} d^{\dagger}_{m}
d^{\dagger}_{m_1} d_{m_1} - \frac{x^2}{2} \sum_{m_1 m_2} d^{\dagger}_{m}
d^{\dagger}_{m_1} d_{m_1}d^{\dagger}_{m_2} d_{m_2} + ...
=
\nonumber\\
= d^{\dagger}_m \left(1 + 2x \hat{N}_d \right)^{1/2}, 
\end{eqnarray}
where the single parameter $x> 0$ determines the strength of 
the nonlinear effects,
i.e., the problem reduces to the harmonic oscillator 
with linear coupling when $x \rightarrow 0$.
%
%
Nonzero values of $x$ that we will consider here lead to
a number of nonlinear effects that results in 
enhancement of the excitation cross sections for the double Giant 
resonances.

Expression (\ref{E1}), together with the standard coupled-channel
formalism presented in the previous section, composes our nonlinear
model. Let us mention its advantageous points:

(i) The model takes into account the nonlinear effects (not all of
them by at least part of them), nonlinearity is governed by a single
parameter

(ii) The model is exactly soluble beyond the perturbative theory

(iii) Its results reproduce correctly the magnitude of the 
{\it enhancement factor} for the Double GDR cross sections 
and its dependence on bombarding energy

\section{Algebraic solution: SU(1,1) dynamics}

To solve the highly nonlinear problem (\ref{U}) with (\ref{BASIC}) and
(\ref{E1}) we introduce the triad of operators
$D^+$,  $D^-$ and $D^0$ that are given by  
\begin{eqnarray} \label{SU11}
D^- = \sqrt{ k + \frac{1}{2}\hat{N}_d} (d_{+1} - d_{-1}  ), \quad
D^+ = \frac{1}{2^{1/2}} (d^+_{+1} - d^+_{-1}  )\sqrt{ 2k + \hat{N}_d},
\nonumber\\
D^0 = \frac{1}{4} 
\left[ (d^+_{+1} -  d^+_{-1} )(d_{+1} - d_{-1} ) + 2(2k+\hat{N}_d)
\right], 
\qquad
\end{eqnarray}
where $D^+$ the Hermitean conjugate $D^+ = \left(D^-\right)^{\dagger}$
with $ N_d \equiv  d^+_{+1} d_{+1} + d^+_{-1} d_{-1}$ and 
$k\equiv(4x)^{-1}$. It is easy to check
that they
obey the commutation relations for the {\it noncompact} 
SU(1,1) algebra 
\begin{equation} \label{ALGEBRA}
\left[ D^- , D^0 \right] = D^-, \quad \left[ D^+ , D^0 \right] = -D^+ , 
\quad \left[ D^- , D^+ \right] = 2 D^0.
\end{equation}
i.e., Eqs.(\ref{SU11}) can be viewed a verision of the Holstein-Primakoff
realization of SU(1,1) algebra. The parameter $k=1/(4x)$ is related
to the Casimir operator of SU(1,1).
 
The dynamics of the system, in the transverse approximation,
can be expressed now in terms of the operators $D^+$,  $D^-$ and $D^0$ 
(\ref{SU11}) only.
The interaction-picture evolution equation (\ref{U}) 
and its formal exact solution, the time-ordered exponential 
(see, e.g., \cite{KIRZHNITS})
than read 
\begin{eqnarray} \label{INTERACTION}
i \frac{d} {d t}U_I(t) = \left[ 
\frac{v_1(t)}{\sqrt{k}} e^{i \omega t}  D^{\dagger} + 
\frac{ v_1(t)}{\sqrt{k}} e^{- i \omega t} D^- 
\right] U_I(t),  
\quad
\nonumber\\
U_I(t) = T 
exp \left( -i \int\limits_{-\infty}^{t} d t' V_I(t') \right) 
\end{eqnarray}
where the commutation relation $[ \hat{N}_d, D^{\pm} ] = \pm D^{\pm}$
and the algebra (\ref{ALGEBRA}) 
has been used in (\ref{U}),(\ref{BASIC}) and (\ref{E1}).

From purely mathematical viewpoint,
the problem described by the last equation
drops into the universality class of the systems
with SU(1,1) dynamics that can be analyzed by means of generalized 
coherent states \cite{PERELOMOV},\cite{PP} for the SU(1,1) algebra. 
(For other algebraic approaches to scattering problems,  
see Refs.\cite{GWI}, \cite{AGI}).

Due to closure of the commutation relations between the operators 
$D^{+},D^{-}$ and $D^0$ that enter the exponential in 
Eq.(\ref{INTERACTION}),
the time-ordered exponential can be represented in another equivalent form
that involve ordinary operator exponentials only 
(see, e.g., \cite{KLEINERT}):
\begin{equation}  \label{DECOMPOSITION}
U_I(t) = 
exp \left[ \frac{\alpha(t)}{\sqrt{k}} D^+ \right] 
exp \left[ \left[ log \left(1 - \frac{|\alpha(t)|^2}{k}\right) -i\phi(t) 
\right] D^0 \right]
exp \left[- \frac{\alpha^{*}(t)}{\sqrt{k}} D^- \right] 
\end{equation}
and
some time-dependent complex number $\alpha(t)$ 
(star means complex conjugation) and real number $\phi(t)$ (phase)
\cite{PERELOMOV}.
The  unknown functions $\alpha(t)$ and $\phi(t)$ can be found from 
simple differential equations 
which relate them 
to the 
function $v_1(t)$ in the Hamiltonian $H(t)$. 
These equations 
can be 
restored after substituting the right hand side of Eq.(\ref{DECOMPOSITION})
into the left hand side of the Schr\"odinger equation for the operator
$U_I(t)$ (\ref{INTERACTION}) and collecting the terms which have the same
operator structure. 
Proceeding this way, 
we obtain, after some algebraic manipulations, 
from (\ref{INTERACTION}) with
using the commutation relations (\ref{ALGEBRA}) and applying 
the known formula 
formula for the operator exponentials
\begin{displaymath}
e^{\hat Y} \hat X e^{-\hat Y} = \hat X + \left[\hat Y,  \hat X
\right] +
\frac{1}{2!} \left[\hat Y,  \left[\hat Y, \hat X
\right] \right] + ...
\end{displaymath}
the following Riccati-type equation for the complex amplitude $\alpha$:
\begin{equation}   \label{RICCATI}
i \frac{d} {d t} \alpha = v_1(t) e^{i \omega t} +
4 x v_1(t) e^{-i \omega t} \alpha^2.
\end{equation}
The phase $\phi(t)$ is given by a simple integral
\begin{displaymath}
\phi(t)= (2/k)\int\limits_{-\infty}^t dt_1 Re( v_1(t_1) 
\alpha(t_1) e^{-i \omega t_1}).
\end{displaymath}
In fact, $\phi(t)$ does not contribute to $W_N$ and $\sigma_N$.
The simple nonlinear equation (\ref{RICCATI}) 
accounts for all orders of 
quantum perturbation theory for the
problem Eqs.(\ref{U}),(\ref{BASIC}),(\ref{INTERACTION}). 
It is also seen from Eq.(\ref{DECOMPOSITION})
that unitarity is automatically preserved within present 
formalism ($U_I^{\dagger}=U_I^{-1}$).

The expression for the amplitudes $a_N(t)$ which we are interesting in 
follows from (\ref{DECOMPOSITION}) immediately after projection of the
state 
\begin{displaymath}
|\psi(t)\rangle = U_I(t) |0 \rangle
\end{displaymath}
onto the intrinsic states with definite number
of GDR phonons, $N$. 
From Eq.(\ref{DECOMPOSITION}), we have 
\begin{displaymath}
U_I(t) | 0 \rangle = 
e^{-i k \phi(t)} \left( 1 - \frac{|\alpha(t)|^2}{ k}  \right)^{k}
exp \left[ \frac{\alpha(t)}{\sqrt{k}} D^+ \right] | 0 \rangle. 
\end{displaymath}
From the last equation, 
we obtain 
the final expression for the amplitude $|a_{N}(\infty)|$ 
and the excitation probabilities for 
the excited states with
$N$ phonons:
\begin{eqnarray} \label{RESULT}
\quad  W_{N} = | a_{N}(\infty) |^2 , \qquad 
\qquad \qquad \qquad \qquad
\nonumber\\
| a_{N}(\infty) | = 
\left( 1 - 4x|\bar{\alpha}(x)|^2  \right)^{\frac{1}{4x}} 
\left( 
\frac{ \Gamma(\frac{1}{2x}+N) }{ N! \Gamma(\frac{1}{2x})} 
\right)
^{1/2} 
\left( 4x|\bar{\alpha}(x)|^2\right)^{N/2}
\end{eqnarray}
Here, the quantity 
$\bar{\alpha}(x)$
is the asymptotic solution to
the Riccati equation (\ref{RICCATI}) at $t \rightarrow \infty$
subject to the initial condition $\alpha(- \infty) = 0$.
The combination $x^{1/2}|\bar{\alpha}(x)|$ in (\ref{RESULT}) 
can be viewed as 
a ``special function'' of the 
two parameters, $x^{1/2}F/\omega$ 
and the adiabaticity parameter
$\frac{\omega b}{v \gamma}$. It can be easily tabulated by solving
(\ref{RICCATI}).  
The cross sections are then obtained from the usual formula (\ref{Pn})
with using (\ref{RESULT}).

The harmonic limit of these results corresponds to the case
$x \rightarrow 0$, 
when the nonlinearity disappears in the transition operators (\ref{E1})
and the coupling to electromagnetic field becomes linear. 
Then at $x \rightarrow 0$ 
the last nonlinear term  drops from the equation (\ref{RICCATI}),
and the amplitude is reduced to its harmonic value
\begin{displaymath}
|\bar{\alpha}(x)|\rightarrow|\alpha^{harm}_{\pm1}|=
\left\vert -i \int\limits_{-\infty}^{\infty}
v_1(t)  e^{i \omega t} dt \right\vert  = 2 \frac{w}{\omega} 
\left(\frac{\omega b}{v \gamma}\right)^2 
K_1\left(\frac{\omega b}{v \gamma}\right)
\end{displaymath}
that is given by the modified Bessel function \cite{WA79},\cite{EMLING}.
The expression for the probabilities $W$ (\ref{RESULT})
reduces at $k \rightarrow \infty$ to the 
Poisson formula (\ref{POISSON}), 
thus the harmonic results \cite{WA79},\cite{EMLING} are restored.


At nonzero nonlinearity $x>0$, the excitation 
probabilities $W_{N}$ (\ref{RESULT}) 
for multiple GDR ($N>1$) turn out to be enhanced as compared to their 
values in the harmonic limit $W^{harm}_{N}$,
as illustrated in Fig.1. 
\begin{figure}[t1]
\rule{5cm}{0.2mm}\hfill\rule{5cm}{0.2mm}
\vskip 3.0cm
\psfig{figure=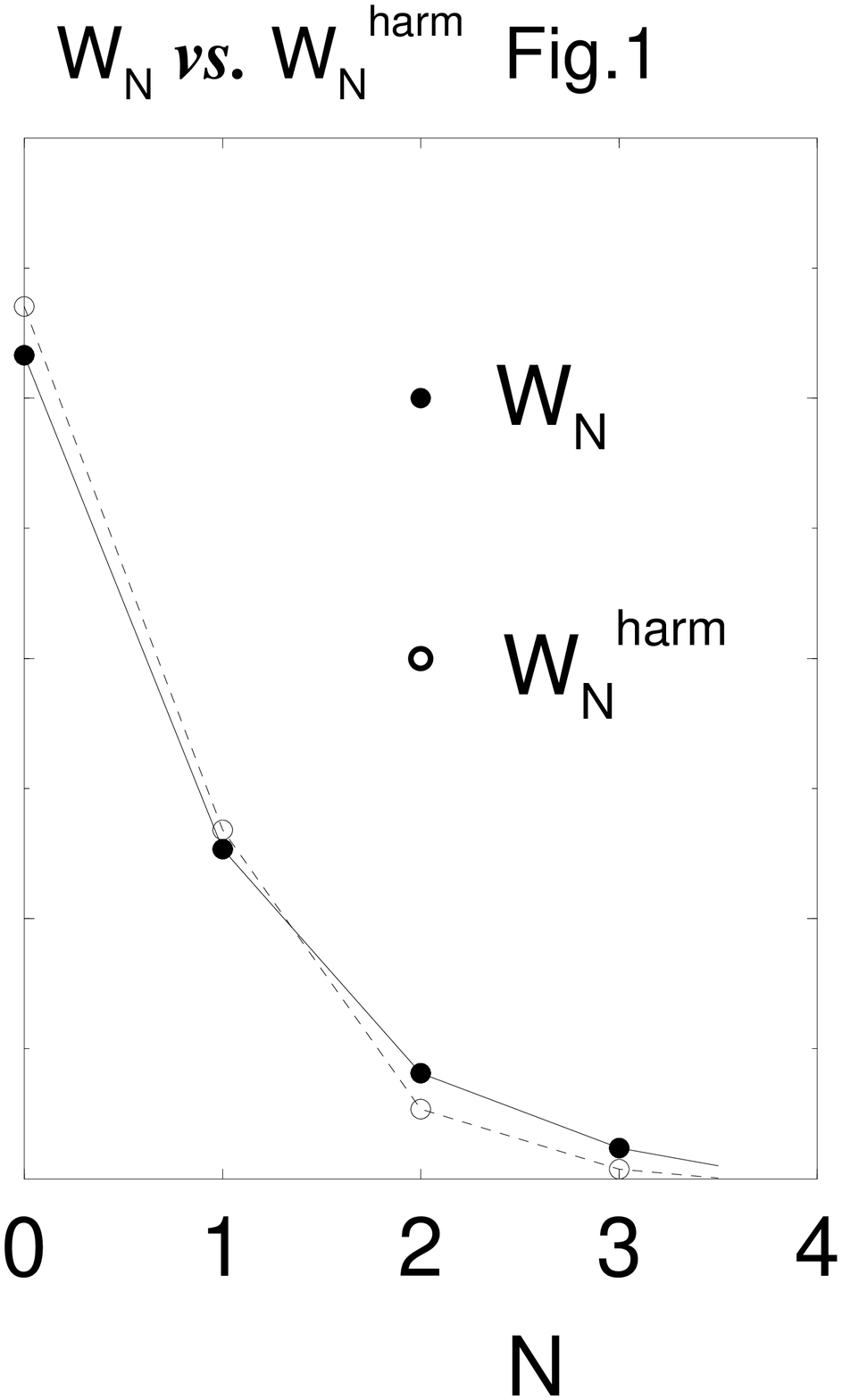,height=2.0in}
\rule{5cm}{0.2mm}\hfill\rule{5cm}{0.2mm}
\caption{
Effect of the nonlinearity on the 
$N$-phonon excitation probabilities(schematic plot).
Nonlinear model values $W_{N}$
(dark circles connected by the solid line) as compared to the 
harmonic model (Poisson) results $W^{harm}_{N}$
(empty circles connected by the dashed line). 
\label{fig:F1}}
\end{figure}

The deviation in the excitation probabilities 
for the $N$-phonon states $W_N$ (\ref{RESULT})
from their harmonic values $W^{harm}_{N}$ (\ref{POISSON})
(the {\it enhancement factor})
is given by the ratio  
\begin{equation}   \label{ENHANCEMENT}
\frac{W_N}{ W^{harm}_{N} } = 
\frac{\Gamma(\frac{1}{2x}+N)}{\Gamma(\frac{1}{2x})\left(\frac{1}{2x}\right)^N}
\frac{( 1 - 4x|\bar{\alpha}(x)|^2)^{\frac{1}{2x}}}
{e^{-2 |\alpha^{harm}_{1}|^2}} 
\frac{ | \bar{\alpha}(x) |^{2N}}{ | \alpha^{harm}_{1} |^{2N} } \quad \geq 1.
\end{equation}

The following ratio is especially representative  
(see (\ref{POISSON}),(\ref{RESULT})) for the two-phonon case
\begin{equation} \label{R2}
R_2 \quad =  
\quad \frac{ W_{2} \quad / \quad W_{1}}
{ W^{harm}_{2} \quad / \quad W^{harm}_{1}}
= \quad (1 + 2x)
\frac{ | \bar{\alpha}(x) |^2}{ | \alpha^{harm}_{1} |^2}.
\end{equation}
The first factor in this expression, 
$1+2x$, 
results
from the kinematic enhancement of the transition probabilities
due to nonlinear effects
considered here. 
The last factor in (\ref{ENHANCEMENT}),(\ref{R2}) results from 
dynamical effects caused by linearity which are incorporated in the
asymptotic solution of the nonlinear equation (\ref{RICCATI}).
Unlike the first factor that
depends on $x$ only, the second one depends on the bombarding
energy and it gives rise to additional enhancement in
low energy domain. The same is valid for the emnhancement factor
in the cross sections, $r_2$, that is given by the 
formula 
\begin{equation} \label{PSI}
r_2 = \frac{\sigma_2}{\sigma^{harm}_{2}} =
\frac{ 2 \pi \int\limits_{b_{gr}}^{\infty} b W_2 db}
{  2 \pi \int\limits_{b_{gr}}^{\infty} b W^{harm}_2 db}.
\end{equation}

\section{Results and discussion}

The interesting feature of these results is that the enhancement
factor  (\ref{ENHANCEMENT}),(\ref{R2}) is much more sensitive
to the bombarding energy than to the parameters of the spectator
partner. 
This is just what has been observed in experiments: the values of $r_2^{exp}$
found for DGDR in $^{208}Pb$ projectile using different targets
$^{120}Sn$, $^{165}Ho$, $^{208}Pb$,  $^{238}U$ \cite{B-208-Pb-1.33} 
are very close to each other and 
they correlate, within the error bars, with the value
\begin{equation}  \label{r2-208Pb}
r_2 (^{208}Pb) \simeq 1.33 \quad , \qquad high \quad energy, \quad
\gamma\simeq 1.7 
\end{equation}
(bombarding energy $\varepsilon \simeq 640$ Mev/per nucleon).

The same picture was found in experiments on Coulomb desintegration 
of $^{197}Au$ target using various projectiles 
$^{20}Ne, ^{86}Kr, ^{197}Au,  ^{209}Bi$ \cite{AU-EXP-Au}.
Also, the similar conclusion of nearly constant value of $r_2$ 
in $^{208}Pb$ target with various projectiles
has been made in work \cite{B-208-Pb-LOW} 
but for low bombarding energy $\varepsilon \simeq 60-100$ Mev/per nucleon.
In this case,
\begin{equation}  \label{r2-208Pb-low}
r_2 (^{208}Pb) \simeq 2 \quad , \qquad low \quad energy, \quad
\gamma\simeq 1.06-1.10. 
\end{equation}

According to the results of calculation within present nonlinear model, 
these enhancement factors would correspond to the nonlinearity
parameter $x$ equal to 
\begin{displaymath}
x(^{208}Pb) \simeq 0.16-0.20
\end{displaymath}
i.e., to a reasonably small nonlinearity.

Below, we present the exact results for the cross sections
calculated according to Eqs.(\ref{Pn}), and (\ref{RESULT}),(\ref{PSI}) 
and with solving Eq.(\ref{RICCATI}) numerically.

\begin{figure}[t2]
\rule{5cm}{0.2mm}\hfill\rule{5cm}{0.2mm}
\vskip 3.0cm
\psfig{figure=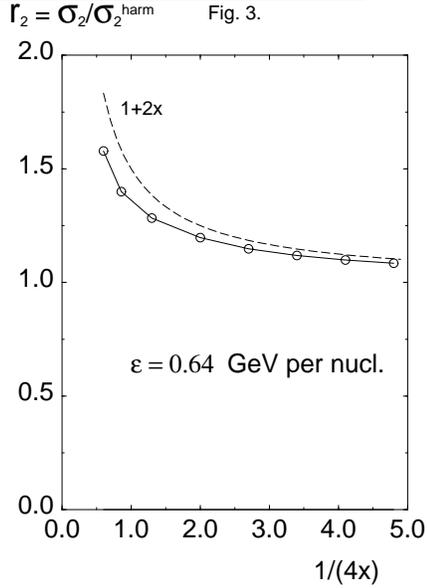,height=2.0in}
\rule{5cm}{0.2mm}\hfill\rule{5cm}{0.2mm}
\caption{
Enhancement factor $r_2 = \sigma_2/\sigma_2^{harm}$
for the Double GDR excitation in 
$^{208}Pb + ^{208}Pb$ process at bombarding energy $\varepsilon=
640 MeV$/per nucleon as a function of the inverse nonlinearity
parameter $k=\frac{1}{4x}$ (symbols, solid
curve is to guide the eye).
Dashed curve corresponds to the scaling $r_2 \simeq (1+2x)=\frac{2 k +1}{2k}$.
\label{fig:F2}}
\end{figure}

The dependence of the enhancement factor $r_2=\sigma_2/\sigma^{harm}_2$ 
(\ref{PSI})
for the DGDR excitation on the strength of
the nonlinearity $x$ is shown on Fig. 1. for the process   
$^{208}Pb + ^{208}Pb$ (we use the case of bombarding energy 
$\varepsilon=0.64 GeV/$per nucleon). It is seen that the enhancement factor 
drops to unity at big values of $k$ (harmonic limit)
and grows at stronger nonlinearity. The scaling value of $r_2$
is also shown for comparison.

In view of the difference between the enhancement factor values 
at high and low energies (Eqs.(\ref{r2-208Pb},\ref{r2-208Pb-low})),
it is interesting to trace energy dependence of the enhancement factor.

Deviations in $r_2$ from the straight line $1+2x$ occur at both low and high
energies. At $\gamma \rightarrow 1$, adiabatic approximation 
is valid, and this yields to
$|\alpha| 
> |\alpha^{harm}_{1}|$.
Thus, $R_2 > 1+2x$ and  $r_2 > 1+2x$.
By contrast, at higher energies, the dynamical nonlinear effects 
tend to reduce the magnitude 
of $|\alpha|$, thus $|\alpha|/|\alpha^{harm}_{1}| <1$,
and $R_2 < 1+2x$.
To sum up, the enhancement factor for the 
DGDR excitation cross section, $r_2 = \sigma_2 / \sigma^{harm}_2$ drops 
from $2-2.5$ (for low bombarding energies $\varepsilon \sim 100 MeV $ per 
nucleon) to
$1.2-1.3$ (for  $\varepsilon \sim 640-700 MeV$ per nucleon) 
while fixed value of nonlinearity 
$x$ is used. 

On Fig.2., we plotted the value of the enhancement factor 
calculated numerically for the case
of $^{208}Pb+^{208}Pb$ process. The magnitude of nonlinearity is
kept fixed  
$x = 0.19$.

It is seen that reasonably small nonlinearity reproduces correctly
the observable value of the enhancement factor and its energy dependence. 
These results are is in correspondence with the experimentally 
observed trends and with 
microscopic models based on Axel-Brink concept \cite{hot-phonon1}.

\begin{figure}[t3]
\rule{5cm}{0.2mm}\hfill\rule{5cm}{0.2mm}
\vskip 1.0cm
\psfig{figure=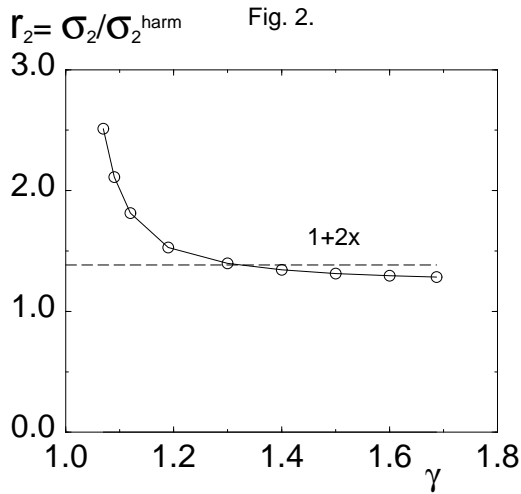,height=2.5in}
\rule{5cm}{0.2mm}\hfill\rule{5cm}{0.2mm}
\caption{
Enhancement factor $r_2 = \sigma_2/\sigma_2^{harm}$
for the Double GDR excitation in 
$^{208}Pb + ^{208}Pb$ process as a function of relativistic factor 
$\gamma$ (symbols, solid curve is to guide the eye). 
The value of the nonlinear parameter $x$ is kept to be equal 
to $x=0.19$. The scaling value (constant $1+2x=\frac{2 k +1}{2k}$) 
is shown by dashed curve.
\label{fig:F3}}
\end{figure}

To conclude, we presented here a simple model that accounts for the 
nonlinear effects in the transition probabilities for the excitation 
of multi-phonon Giant Dipole Resonances in Coulomb excitation 
via relativistic heavy ion collisions. The model is based on the 
group theoretical properties of the boson operators. It allows 
to construct the solution for the dynamics of the multi-phonon 
excitation within coupled-channel approach in terms of the 
generalized coherent states of the corresponding algebras.
The 
exactly solvable harmonic phonon model 
appears to be a limiting case of the present model when the 
nonlinearity parameter $x$ goes to zero. 
The main advantages of the limiting harmonic case (unrestricted
multiphonon basis, preservation of unitarity and 
analytical results in nonperturbative domain) remain present in our nonlinear
scheme.
Therefore, the model can be viewed as a natural extension of
the harmonic phonon model to include the nonlinear effects in 
a consistent way while keeping the model solvable.

The probabilities to excite double-phonon
GDR appear to be enhanced by a factor 
$(1 + 2x)(|\bar{\alpha}(x)|^2/|\alpha^{harm}_1|^2$; 
that results in the cross section enhancement factor.
This can be viewed as a hint that the discrepancy 
between the measured cross-sections of double GDR and the harmonic
phonon calculations can be resolved within present nonlinear model
by means of using an appropriate value of the nonlinear parameter 
$x$ for a given nucleus. 
The experimental values of enhancement
of $\sigma_2$ with respect to the harmonic results for the excited 
$^{208}Pb$ nucleus are almost insensitive to the details of the
collision process.
The enhancement factor drops as the bombarding
energy grows. 
This is consistent with the data and 
gives results similar to those recently obtained in a possibly
different context, with a theory based on the concept of fluctuations
(damping) and the Brink-Axel mechanism 
\cite{hot-phonon1},\cite{hot-phonon2},\cite{BNW},\cite{CHP-HOT-3}.
It would be certainly worthwhile to establish possible connections
between the two approaches.

\section*{Acknowledgments}
The work has been supported by FAPESP 
(Fundacao de Amparo a Pesquisa do Estado de Sao Paulo).


\end{document}